\documentstyle[11pt,moriond,epsfig]{article}
\def\Journal#1#2#3#4{{#1} {\bf #2}, #3 (#4)}

\def\NPB{{\em Nucl. Phys.} B}
\def\PLB{{\em Phys. Lett.}  B}

\def\PRD{{\em Phys. Rev.} D}
\def\ZPC{{\em Z. Phys.} C}
\def\EPJC{{\em Eur. Phys. Jour.} C}
\def\AP{\em Ann. Phys.}
\def\IJMPA{{\em Int. Jour. of Mod. Phys.} A}

\def\be{\begin{equation}}
\def\ee{\end{equation}}
\def\bea{\begin{eqnarray}}
\def\eea{\end{eqnarray}}

\begin{document}

\vspace*{4cm}
\title {VALUE OF $\alpha_s$ AND HIGH TWISTS \\
FROM COMBINED ANALYSIS OF $e-\mu$ DIS DATA}

\author{ S. ALEKHIN}

\address{Institute for High Energy Physics, Protvino, 142284, Russia}

\maketitle\abstracts
{We perform a NLO QCD analysis of the combined SLAC-BCDMS-NMC-E665-H1-ZEUS 
data on inclusive deep inelastic 
cross section. Particular attention was paid to the extraction of 
strong coupling constant $\alpha_s$
and high twist (HT) contribution to the structure functions $F_2$ and $F_L$.  
It was shown that at small and moderate 
x there is a visible dependence of the extracted values of HT 
contribution to $F_2$ on the 
QCD renormalization scale, which indicates that in this region 
extracted HT can absorb NNLO QCD corrections. At larger x the dependence 
of HT on the renormalization scale is negligible and the influence 
of NNLO correction on their values should be less significant.
The value of $\alpha_s(M_Z)=0.1159\pm0.0031$ (total) is obtained, where
the error includes statistical, systematical and theoretical uncertainties.}

The problem of strong coupling constant
determination from charged leptons
deep inelastic scattering (DIS) data was widely discussed 
recent years. These data are very precise (at the level of $O(1\%)$)
and the theoretical uncertainties of the analysis are relatively small, 
which allows for to determine the value of $\alpha_s(M_Z)$ with the 
precision of $O(0.001)$. At the same time there is statistically 
significant discrepancy between the value of $\alpha_s$ obtained
in the analysis \cite{VM} of combined SLAC-BCDMS proton-deuterium
data \cite{SLAC,BCDMS} and the 
results of experiments performed at LEP \cite{BET}.
This discrepancy can be considered as an indication on new physics 
beyond Standard Model \cite{GF}. Meanwhile, 
as it became clear in the earliest QCD analysis of DIS data
the value of $\alpha_s$ is strongly correlated with the value of 
possible high twist (HT) contribution to the structure function $F_2$
\cite{AB}. This correlation
 makes the
separation of log-like and power-like contributions to the scaling 
violation unstable with respect to various assumptions made in 
the analysis. In particular, as it was shown recently \cite{AL98},
the results of nonsinglet SLAC-BCDMS proton-deuterium data analysis are 
sensitive to the procedures used to handle systematic errors on the data.
The central value of $\alpha_s(M_Z)=0.1180\pm0.0017$ (stat.+syst.),
as obtained in the analysis of Ref.\cite{AL98}
with the complete account of point-to-point 
correlations due to systematic errors,
is significantly larger than the results of Ref.\cite{VM} and is 
compatible with the LEP measurements and world average.
In the extended version of this analysis with addition of the NMC 
proton-deuterium data \cite{NMC}
and account of HT contribution to the structure function $F_L$
the value of  $\alpha_s(M_Z)=0.1170\pm0.0021$ (stat.+syst.) 
was obtained \cite{AL99}. In this talk we describe 
the effect of further extension of the analysed data set  
on the value of $\alpha_s(M_Z)$.

 The analysis was performed in NLO QCD approximation
with fixed number of evolved fermion distributions $(N_f=3)$;
momentum sum rule (MSR) and fermion sum rule for valence 
quarks were used to decrease the number of fitted parameters;
HT contributions to the structure functions $F_2$ and $F_L$
were parametrized in additive form;
target mass correction \cite{TMC} 
was accounted for in the fitted cross section formula;  
account of systematic errors was performed through 
the covariance matrix approach.
More detailed description of the ansatz
 can be found in the earlier papers 
\cite{AL98,AL99,AL96}. The milestone results are 
given in Table 1. From the top to bottom rows the precision and 
reliability of the $\alpha_s$ improves due to more data included 
in the analysis and more corrections applied. The cut $Q^2>2.5$~GeV$^2$
was imposed because 
without this cut the total 
error is dominated by the uncertainty in QCD renormalization scale 
(see fourth row of Table 1) and it is meaningless to suppress
the experimental error with a risk to encounter 
an unexpected small $Q^2$ effects. Our final value of $\alpha_s(M_Z)$
is given in the last row of Table 1 and its various theoretical 
uncertainties -- in Table 2. With the account of the shifts due to 
this uncertainties one can obtain 
\begin{displaymath}  
\alpha_s(M_Z)=0.1159\pm0.0031 {\rm (total)},
\end{displaymath}
that is compatible with world average. 
In the analysis of Ref.\cite{VOGT} it was obtained,
that in the QCD fit to the world DIS data,
performed without low $Q^2$ cut, MSR is violated.
In order to check this conclusion a test fit without imposing MSR boundary
condition was made. Obtained value of total momentum carried by partons 
is $<x>=0.982\pm0.028$ that is compatible with 1 within errors 
and is in disagreement with the results of Ref.\cite{VOGT} 
($<x>\approx1.08\pm0.02$ for the cut $Q^2\ge 3$~GeV$^2$). 

\begin{table}[h]
\begin{center}
\caption{The values of $\alpha_s(M_Z)$ obtained in the 
analysis of different data sets. NDP is total  
number of data points,
experimental error (exp) includes
statistical and systematical uncertainties, (RS) is the shift due to 
the change of QCD renormalization scale from $Q^2$ to $4Q^2$.}
\begin{tabular}{|c|c|c|} \hline
{\bf Analysed data set} & NDP& $\alpha_s(M_Z)$ \\ \hline
SLAC-BCDMS-NMC $(0.3\le x \le 0.75)$ & 1243 &$0.1170\pm0.0021$(exp) \\ \hline
SLAC-BCDMS-NMC $(0.3\le x \le 0.75,$ & 1243 &  \\ 
Fermi motion correction \cite {FERMI} (FMC) on)& &$0.1190\pm0.0020$(exp) \\ \hline
SLAC-BCDMS-NMC $(x\ge 0.3,$ FMC on) & 1348 &$0.1197\pm0.0019$(exp) \\ \hline 
SLAC-BCDMS-NMC (no x-cut, FMC on) & 2541 &$0.1190\pm0.0012({\rm exp})\pm$ \\ 
  & &$+0.0028$(RS) \\ \hline 
SLAC-BCDMS-NMC  & 2083 &\\ 
(no x-cut, $Q^2>2.5~$GeV$^2$, FMC on) & &$0.1170\pm0.0019$(exp) \\ \hline
SLAC-BCDMS-NMC-E665\cite{E665}-H1\cite{H1}-ZEUS\cite{ZEUS} & 2512 & \\ 
(no x-cut, $Q^2>2.5~$GeV$^2$, FMC on)& &$0.1166\pm0.0016$ (exp) \\ \hline
\end{tabular}
\end{center}
\end{table}

\begin{table}[h]
\begin{center}
\caption{The effect of various theoretical uncertainties on the 
value of $\alpha_s(M_Z)$.} 
\begin{tabular}{|c|c|} \hline
{\bf Source of uncertainty} & $\Delta\alpha_s(M_Z)$ \\ \hline
QCD renormalization scale variation from $1/4Q^2$ to $4Q^2$ & $-0.0022,+0.0028$
\\ \hline
The change of matching scale of heavy & \\ 
quark threshold from $m_Q^2$ to $6.5m_Q^2$(c.f. Ref.\cite{BLUM}) & --0.0020 \\ \hline
The change of c-quark mass on $\pm0.25$~GeV & $\pm0.0002$ \\ \hline
The change of strange sea suppression factor on $\pm0.1$ & $\pm0.0001$ \\ \hline
\end{tabular}
\end{center}
\end{table}

\begin{figure}[t]
\centerline{\psfig{file=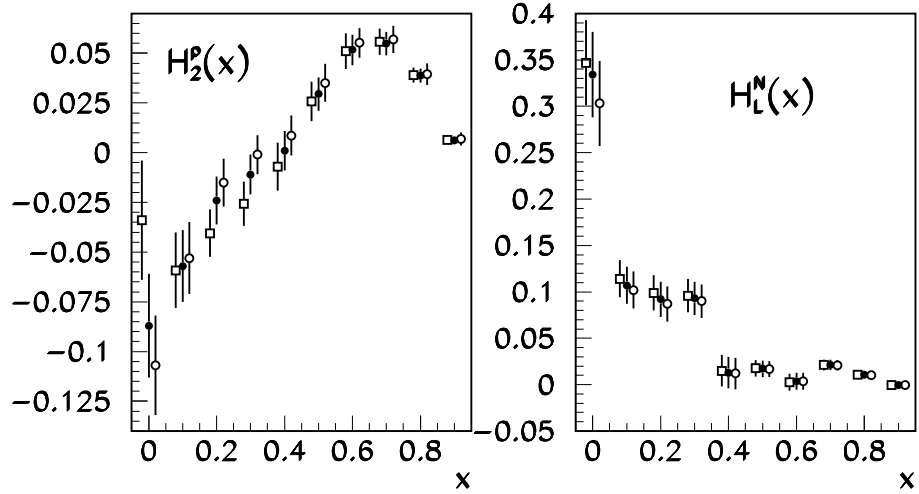,height=6.5cm}}
\caption{The fitted values of $H_2^p(x)$ and $H_L^N(x)$ for 
different choices of the QCD renormalization scale $\mu$.
Full circles correspond to $\mu=Q^2$, empty circles -- 
to $\mu=4Q^2$, squares -- to $\mu=Q^2/4$}
\end{figure}

The fitted values of HT contributions to the proton 
structure function $F_2$ and nucleon structure function $F_L$
for different choices of the QCD renormalization scale $\mu$
are given in Fig. 1. (It was adopted in the analysis that 
HT contributions to the proton and neutron structure functions
$F_L$ are equal, since data cannot discriminate between them).
The visible dependence of $H_2(x)$ on $\mu$ at small x can be considered 
as an indication 
that in this region HT contribution to $F_2$
 absorbs NNLO QCD 
corrections. The indication on interplay between HT contribution 
to the structure function $F_3$
and NNLO corrections was obtained in the 
NLO QCD analysis of neutrino data~\cite{ALKATFR};
this interplay was also directly 
demonstrated in the earlier NNLO QCD analysis of Ref.\cite{KAT}.
At the same time for $H_2(x)$ at largest x and 
for $H_L(x)$  this dependence is not so significant. 
It is worth to note that
in the fit with simultaneous extraction of $\alpha_s$ and HT contribution 
the dependence of the $\alpha_s$ value on $\mu$ is   
weaker, than in the fit with HT fixed 
since HT are readjusted with the change of $\mu$;
see in this  connection Fig. 3 of Ref.\cite{ALMOR99}. 
In particular, due to this effect our 
value of the RS uncertainty 
on $\alpha_s(M_Z)$ is less than it was obtained in 
Ref.\cite {VOGT}. At the same time due to the large correlation 
between $\alpha_s$ and HT contribution to $F_2$ the $\alpha_s$ error
increases as compared to the fit with HT fixed and hence one can say 
that some part of the RS error on $\alpha_s$ obtained in the fit 
with HT released is included in the total experimental error.

I am indebted to G.Korchemsky and L.Mankiewicz for useful discussions.


\section*{References}
\begin{thebibliography}{99}

\bibitem{VM}M. Virchaux and A. Milsztajn, \Journal{\PLB}{274}{221}{1992}.

\bibitem{SLAC} L.W. Whitlow et al., \Journal{\PLB}{282}{475}{1992}.

\bibitem{BCDMS}BCDMS collaboration, A.C. Benvenuti et al., 
\Journal{\PLB}{223}{485}{1989};\\
BCDMS collaboration, A.C. Benvenuti et al., \Journal{\PLB}{237}{592}{1990}.

\bibitem {BET} S. Bethke, Preprint PITHA-98-43, Sep. 1998; hep-ph/9812026.

\bibitem{GF} M. Shifman, \Journal{\IJMPA}{11}{3195}{1996}.

\bibitem{AB} L.F. Abbott, W.B. Atwood, and R.M. Barnett, 
\Journal{\PRD}{22}{582}{1980}.

\bibitem{AL98} S.Alekhin, \Journal{\PRD}{59}{114016}{1999}

\bibitem{NMC}
 NM collaboration, M. Arneodo et al., \Journal{\NPB}{483}{3}{1997}.

\bibitem{AL99} S.Alekhin, hep-ph/9902241, Eur. Phys. Jour. C in print.

\bibitem{TMC}
 H. Georgi and H.D. Politzer, \Journal{\PRD}{14}{1829}{1976}.

\bibitem{AL96} S.Alekhin, \Journal{\EPJC}{10}{395}{1999}.

\bibitem{FERMI} G.B. West,  \Journal{\AP}{74}{464}{1972}.

\bibitem{E665} E665 collaboration, M.R.Adams et al., 
\Journal{\PRD}{54}{3006}{1996}.

\bibitem{H1} H1 collaboration, S. Aid et al., \Journal{\NPB}{470}{3}{1996}.

\bibitem{ZEUS} ZEUS collaboration, M. Derrick et al.,
\Journal{\ZPC}{72}{399}{1996}.

\bibitem{BLUM} J. Bl\"{u}mlein and W.L. van Neerven, 
\Journal{\PLB}{450}{417}{1999}.

\bibitem{VOGT} A. Vogt, hep-ph/9906337, to appear in proceedings
of DIS-99.

\bibitem{ALKATFR} S. Alekhin and A. Kataev, hep-ph/9908349, to appear
in proceedings of Nucleon-99.

\bibitem{KAT} A. Kataev et al., \Journal{\PLB}{417}{374}{1998}.  

\bibitem {ALMOR99} S.Alekhin, hep-ph/9907350, to appear in proceedings 
of QCD Moriond-99.

\end{thebibliography}
\end{document}